\documentclass[onecolumn,showpacs,preprintnumbers,amsmath,amssymb]{revtex4}
\usepackage{graphicx}% Include figure files
\usepackage{dcolumn}% Align table columns on decimal point
\usepackage{bm}% bold math
\usepackage{epstopdf}% allow eps to be viewed

\begin{document}

\preprint{}

\title{Input spectrum for heterodyne detection in advanced gravitational wave detectors without non-stationary shot-noise}% Force line breaks with \\

\author{Peter T. Beyersdorf}
 \altaffiliation[Also at  ]{Ginzton Laboratory, Stanford University}%Lines break automatically or can be forced with \\
%\author{Secondary Author}%
 %\email{Second.Author@institution.edu}
\affiliation{%
Physics Department, San Jose State University}%

%\author{Charlie Author}
% \homepage{http://www.Second.institution.edu/~Charlie.Author}
%\affiliation{
%Second institution and/or address\\
%This line break forced% with \\
%}%

\date{\today}% It is always \today, today,
             %  but any date may be explicitly specified

\begin{abstract}
Future interferometric gravitational wave detectors will make use of the coupling between shot noise and radiation pressure noise that produces a squeezed output for the quantum noise at the dark-port of the interferometer allowing these interferometers to operate with a sensitivity that exceeds the standard quantum limit at certain frequencies.  The ability of the detector to take advantage of this squeezed output state depends in part on the readout system used.  With the phase modulated laser input planned for these interferometers the quantum-noise limited sensitivity is slightly better when homodyne (rather than heterodyne) detection is used.  
We show that by modifying the laser input spectrum for these interferometers, readout with heterodyne detection can be made to have the same quantum-noise limited sensitivity as readout with homodyne detection.
\end{abstract}

% PACS, the Physics and Astronomy
 \maketitle

\section{\label{sec:level1}Introduction}

The prospect of operating interferometric gravitational wave detectors for quantum non-demolition measurements\cite{KLMTV} has lead to investigation into the details of the interferometer read-out scheme \cite{Buonanno2003, Somiya2003}.  Two readout schemes relevant for gravitational wave detection are heterodyne readout (often called RF readout) and homodyne readout (often called DC readout).  The readout scheme converts the field at the output port of the interferometer into a signal.  In heterodyne readout a local oscillator field or fields that have a frequency difference from the field being measured interfere with the output field to produce an intensity that has a component which beats at the frequency difference of the fields.  The phase of the beat note is a measure of the phase of the output field.  In homodyne detection a local oscillator field at the same frequency as the field being measured interferes with the output field to produce constructive or destructive interference depending on the relative phase of the output field.

A principle issue for interferometric gravitational wave detectors  is the difference in the quantum-noise-limited sensitivity of an interferometer with homodyne readout versus heterodyne readout.  Buonanno, Chen and Mavalvala\cite{Buonanno2003} show that an interferometer with homodyne readout has better quantum-noise limited sensitivity than with heterodyne readout.  Although other technical tradeoffs exist between homodyne and heterodyne detection \cite{Fritschel2003}, we show that with a change to the laser input field assumed by Buonanno, Chen and Mavalvala it is possible for an interferometer with heterodyne detection to reach the same quantum-noise-limited sensitivity that can be achieved with  homodyne detection.

\section{Input Spectrum}
\begin{figure}[!h]
\includegraphics[width=2 in]{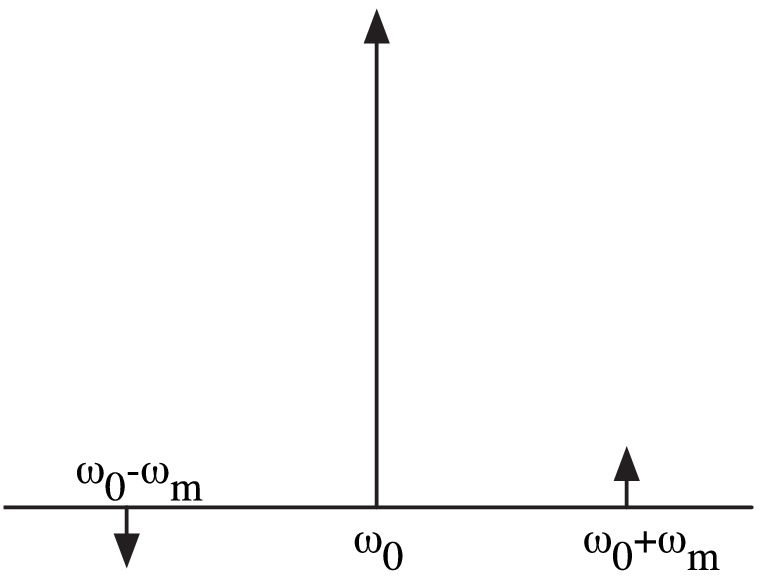}
\includegraphics[width=2 in]{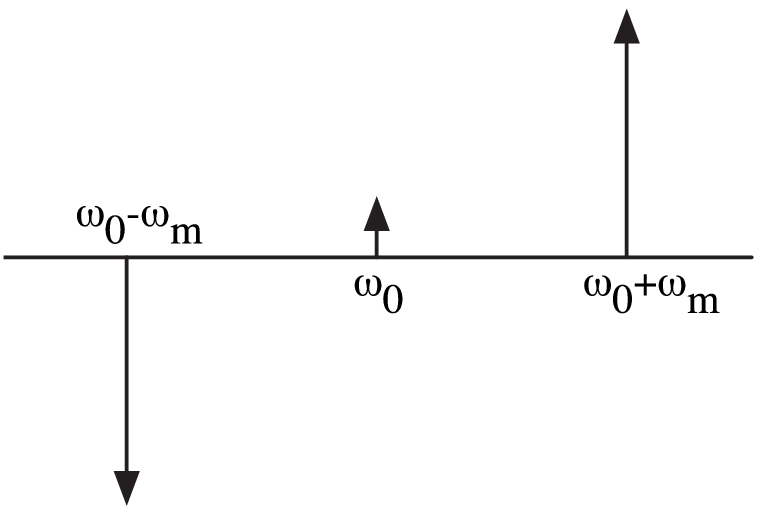}
\caption{\label{fig:inputspectrum}Left: The input spectrum commonly used in gravitational wave interferometers.  It has a high power component at frequency $\omega_{0}$ which resonates in the arms of the interferometer and low power modulation sidebands at $\omega_{0}\pm\omega_{m}$ which transmit to the dark port and are used as a local oscillator for heterodyne detection.  We refer to this as the ``conventional input spectrum''.  Right: A ``suppressed carrier input spectrum'' with a low power component at frequency $\omega_{0}$ which transmit to the dark port to be used as a local oscillator for heterodyne detection, and high power modulation sidebands at  $\omega_{0}\pm\omega_{m}$ which resonate in the arms of the interferometer}
\label{fig:inputspectrum}
\end{figure}

The conventional laser input spectrum for interferometric gravitational wave detectors is shown in the left-hand panel of figure \ref{fig:inputspectrum}.  It consists of a high-power carrier at frequency $\omega_{0}$ which is resonant in the arms and interferes destructively at the dark-port of the interferometer, along with a pair of sidebands at $\omega_{0}\pm \omega_{m}$ that do not resonate in the arms and are made to transmit efficiently to the dark-port of the interferometer via an asymmetry in the arm lengths.  Depending on the implementation there may be additional pairs of sidebands at other frequencies that are used for alignment sensing and control, which are not relevant for the discussion of the readout noise and which we shall ignore in our analysis.  In heterodyne detection one or both of the sidebands at $\omega_{0}\pm \omega_{m}$ are transmitted to the dark port of the interferometer where they interfere with the gravitational wave signal sidebands of the carrier field at frequency $\omega_{0}\pm \Omega_{gw}$, where $\Omega_{gw} \ll \omega_{m}$ is the gravitational wave frequency which is small compared to the modulation frequency.  The sidebands serve as the local oscillator for heterodyne detection of signals in a frequency band centered on the carrier frequency.  With this input laser spectrum  the signal sidebands at frequencies near $\omega_{0}$ are detected, along with the vacuum fluctuations (quantum noise) that have frequencies in a band centered on $\omega_{0}$, as well as the fluctuations that have a frequency in either of the frequency bands centered on $\omega_{0}\pm 2 \omega_{m}$.  Since the spectral region around $\omega_{0}\pm 2 \omega_{m}$ contributes noise when it mixes with the local oscillator fields at $\omega_{0}\pm\omega_{m}$ and is demodulated, but does not contribute any signal, it degrades the quantum-noise-limited sensitivity of the interferometer.  By contrast with homodyne detection a field at the carrier frequency $\omega_{0}$ is used as the local oscillator so that  only the signal and quantum noise in the spectral region around $\omega_{0}$ contribute.  Thus with homodyne detection there is less quantum-noise added to the signal.  

The carrier-suppressed input spectrum shown in the right-hand panel of Figure \ref{fig:inputspectrum} leads to a difference in the way vacuum fluctuations affect the sensitivity.  It consists of a low powered local oscillator at $\omega_{0}$ and two high-powered sidebands at  $\omega_{0}\pm \omega_{m}$.  The sidebands are assumed to be resonant in the arm cavities while the local oscillator at frequency $\omega_{0}$ may or may not be resonant in the arm cavities depending on the implementation.  Like with the conventional input spectrum, the Michelson interferometer's interference condition is set so that the high power components that resonate in the arm cavities interfere destructively at the dark-port while the low-power component that is used as a local oscillator is transmitted to the dark port of the interferometer.  
For comparison to the case of heterodyne readout with the conventional input spectrum or homodyne readout, we shall assume the same total power in the input spectrum is the same,  so that the sidebands of our input spectrum each have half the power of the carrier in the conventional input spectrum (and the low power local oscillator components are assumed to have negligible power).  With the suppressed-carrier input spectrum the local oscillator detects the gravitational wave signal sidebands in the frequency bands centered on $\omega_{0} \pm \omega_{m}$ as well as the vacuum fluctuations at these same frequencies.  Unlike heterodyne detection with the conventional input spectrum, \textit{there is no additional quantum noise from frequency regions that do not contribute to the readout of the gravitational wave signal}.  To analyze the quantum-noise limited sensitivity of a configuration with this supressed-carrier input spectrum we must consider the input-output relations for quantum noise in frequency bands centered on both the upper and lower laser sideband frequencies.   We extend the analysis of the input-output relations for a single frequency range done by Buononno and Chen \cite{Buonanno2001} in which the quantum noise is represented by a 2-element vector and the interferometer is represented by a 2x2 matrix by describing the quantum noise with a 4-element vector and representing the interferometer by a 4x4 matrix.

\section{Input-Output Relations}

We consider vacuum fluctuations at the dark port of the interferometer as the source for the quantum noise that is read out.  Specifically we look at the components of the vacuum fluctuations at frequencies $\omega_{0}\pm\omega_{m}$ which are in-phase and quadrature-phase with the high-power sideband at each of those frequencies.  We  call the amplitude of the  vacuum fluctuations that have a frequency in a band centered on the lower laser sideband frequency $\omega_{0}-\omega_{m}$ and are are in-phase and quadrature phase with this sideband $a_{1}$ and $a_{2}$ respectively. We let $a_{3}$ and $a_{4}$ represent the components of the vacuum fluctuations at frequencies in a band centered on the upper sideband frequency of $\omega_{0}+\omega_{m}$ that are in-phase and quadrature phase respectively relative to this sideband.  With this notation, we consider the fields that contribute to the signal readout by heterodyne detection as a 4-element vector.  The interferoemter can then be described as a $4\times4$ matrix that transforms the (unsqueezed) input quantum noise into a squeezed state at the output.  By considering the action of a gravitational wave induced motion of the arm cavity mirrors on the output fields we relate the quantum-noise after demodulation to an equivalent signal magnitude to determine the quantum-noise limited sensitivity. 

\begin{figure}[!h]
\includegraphics[width=2 in]{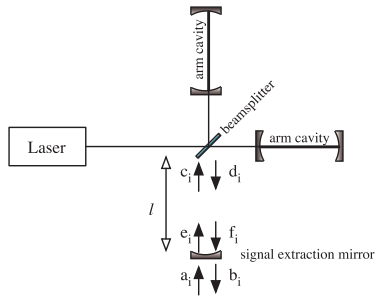}
\caption{\label{fig:interferoemterdiagram}Diagram of a resonant sideband extraction interferometer.  The operating point of the interferometer is set such that the laser power is reflected back towards the laser, and vacuum fluctuations from the dark port $a_{i}$ are reflected to the dark port $b_{i}$}
\label{fig:ifodiagram}
\end{figure}

We start with the input output relations for a conventional interferometer calculated by Kimble et al (eqaution 16 of Ref. \cite{KLMTV}) 
\begin{eqnarray}
d_{1}&=&\Delta d_{1}=c_{1} e^{2 i \beta}\nonumber \\
d_{2}&=&\Delta d_{2}+\sqrt{2 \kappa} \frac{h}{h_{sql}} e^{ i \beta}, \Delta d_{2}=(c_{2}-\kappa c_{1}) e^{2 i \beta}
\label{eq:cd2}
\end{eqnarray} 
with the notation change of $b\to d$ and $a \to c$ to represent the internal fields of the interferometer configuration shown in figure \ref{fig:ifodiagram}.   We extend these equations to account for the additional frequency components of the light circulating in the interfeormeter arms as follows.  First these input-output relations which describe the fields at one frequency of interest (i.e. that of the lower frequency sideband, $\omega_{0}-\omega_{m}$) are used to describe a second frequency of interest (i.e. that of the upper frequency sideband, $\omega_{0}+\omega_{m}$) by making the replacements $d_{1} \to d_{3}$, $c_{1} \to c_{3}$,  $d_{2} \to d_{4}$ and  $c_{2} \to c_{4}$.  Next we add a term which accounts for a cross-coupling between these frequencies bands as follows:  Differential motion of the arm cavity mirrors couples fields out of the interferometer into the dark port with an efficiency proportional to the magnitude of the field in the arms and represented by $\sqrt{\kappa}$.  Analogous to equation [18] of Ref. \cite{KLMTV} we set 
\begin{equation}
\kappa_{\pm}=\frac{(I_{\pm}/I_{sql})2 \gamma^{4}}{\Omega^{2}(\gamma^{2}+\Omega^{2})}
\end{equation}   
Where $I_{\pm}$ is the intensity of the upper and lower sideband, $I_{sql}=\frac{m L^{2} \gamma^{4}}{4 \omega_{0}}$ is the intensity needed to reach the standard quantum limit at the central frequency $\omega_{0}$ for an interferometer with arm lengths L and mirrors suspended as free masses with mass m.  $\Omega$ is the (gravitational wave induced) sideband frequency measured relative to the laser sideband at $\omega_{0}\pm \omega_{m}$.  $2 \gamma\equiv\frac{Tc}{4 L}$ is the arm cavities bandwidth when the mirror that couples light into and out of the cavity has a power reflectivity of T.  The in-phase vacuum fluctuations entering the dark port interfere with the laser power in the arms to produce radiation pressure noise that drives the mirrors differentially by an amount proportional to the interfering laser field $\sqrt{\kappa_{\pm}}$.  Vacuum fluctuations at frequencies close to  the lower sideband frequency $\omega_{0}-\omega_{m}+\Omega$ only produce significant levels of radiation pressure noise by interacting with the lower sideband, and vice versa for the upper sideband.  The coupling of the input field that is in-phase with the laser produces the term $-\kappa c_{1}$ in equation \ref{eq:cd2} by an interaction of the fields that produces a displacement of the mirrors proportional to $\sqrt{\kappa}$ and then a coupling of the laser field to the output through this displacement that is also proportional to $\sqrt{\kappa}$.  Generalizing to the case where laser fields at two different frequencies are resonating in the arms, the displacement of the mirrors is driven by radiation pressure noise from both frequency regions, and is thus proportional to $\sqrt{\kappa_{-}} c_{1}+\sqrt{\kappa_{+}} c_{3}$.  This couples the field $\sqrt{\kappa_{-}}$ to the output port as  $\kappa_{-} c_{1}+\sqrt{\kappa_{-} \kappa_{+}} c_{3}$ and the field $\sqrt{\kappa_{+}}$ to the output port as  $\sqrt{\kappa_{-} \kappa_{+}}  c_{1}+ \kappa_{+} c_{3}$, allowing us to express the input-output relations as 
\begin{eqnarray}
d_{1}&=&\Delta d_{1}=c_{1} e^{2 i \beta}\nonumber \\
d_{2}&=&\Delta d_{2}+\sqrt{2 \kappa_{-}} \frac{h}{h_{sql}} e^{ i \beta}, \Delta d_{2}=(c_{2}-\kappa_{-} c_{1}-\sqrt{\kappa_{-} \kappa_{+}} c_{3}) e^{2 i \beta}\nonumber \\
d_{3}&=&\Delta d_{3}=c_{3} e^{2 i \beta}\nonumber \\
d_{4}&=&\Delta d_{4}+\sqrt{2 \kappa_{+}} \frac{h}{h_{sql}} e^{ i \beta}, \Delta d_{4}=(c_{4}-\kappa_{+} c_{3}-\sqrt{\kappa_{-} \kappa_{+}} c_{1}) e^{2 i \beta}
\label{eq:cd}
\end{eqnarray} 

The relationship between the fields in the interferometer can be expressed by matrices as
%%%%%%%%%%%%Equation Break%%%%%%%%%%%%%%%%%
\begin{equation}
\left[\begin{array}{c}
 c_{1}\\c_{2}\\c_{3}\\c_{4}
\end{array}\right]
=
\left[\begin{array}{cccc}
 \cos\phi&-\sin(\phi)&0&0\\
 \sin\phi&\cos(\phi)&0&0\\
 0&0&\cos\phi&-\sin(\phi)\\
 0&0&\sin\phi&\cos(\phi)\\
\end{array}\right]
\left[\begin{array}{c}
 e_{1}\\e_{2}\\e_{3}\\e_{4}
\end{array}\right]e^{i \Phi} 
\end{equation}
%%%%%%%%%%%%Equation Break%%%%%%%%%%%%%%%%%
\begin{equation}
\left[\begin{array}{c}
 f_{1}\\f_{2}\\f_{3}\\f_{4}
\end{array}\right]
=
\left[\begin{array}{cccc}
 \cos\phi&-\sin(\phi)&0&0\\
 \sin\phi&\cos(\phi)&0&0\\
 0&0&\cos\phi&-\sin(\phi)\\
 0&0&\sin\phi&\cos(\phi)\\
\end{array}\right]
\left[\begin{array}{c}
 d_{1}\\d_{2}\\d_{3}\\d_{4}
\end{array}\right]e^{i \Phi}
\end{equation}
%%%%%%%%%%%%Equation Break%%%%%%%%%%%%%%%%%
\begin{equation}
\left[\begin{array}{c}
 d_{1}\\d_{2}\\d_{3}\\d_{4}
\end{array}\right]
=
\left[\begin{array}{cccc}
 1&0&0&0\\
 -\kappa_{-}&1& -\sqrt{\kappa_{-}\kappa_{+}}&0\\
 0&0           &1&0\\
 -\sqrt{\kappa_{-}\kappa_{+}}&0&-\kappa_{+}&1\\
\end{array}\right]
\left[\begin{array}{c}
 c_{1}\\c_{2}\\c_{3}\\c_{4}
\end{array}\right]e^{2 i \beta}+
\left[\begin{array}{c}
 0\\ \sqrt{2 \kappa_{-}}\\0\\ \sqrt{2 \kappa_{+}}
\end{array}\right]\frac{h}{h_{sql}}e^{i \beta}
\end{equation}
%%%%%%%%%%%%Equation Break%%%%%%%%%%%%%%%%%
\begin{equation}
\mathbf{e}=\rho\mathbf{f} + \tau \mathbf{a}
\end{equation}
%%%%%%%%%%%%Equation Break%%%%%%%%%%%%%%%%%
\begin{equation}
\mathbf{b}=-\rho \mathbf{a} + \tau \mathbf{f}
\end{equation}
%%%%%%%%%%%%Equation Break%%%%%%%%%%%%%%%%%
Where 
\begin{equation}
2\beta\equiv 2\arctan{(\Omega/\gamma)}
\end{equation} is the phase shift acquired by a field at frequency $\Omega$ relative to the sideband frequencies upon reflecting from the arm cavities, $\phi=(\omega_{0}\pm\omega_{m}) l/c$ is the phase gained by the modulation sidebands while traveling one way in the signal cavity, and $\Phi=\Omega l/c$ is the additional phase gained by a (gravitational wave or quantum-noise induced) sideband at frequency $\omega_{0}\pm\omega_{m}+\Omega$.
These relationships allows us to solve for the output field $\mathbf{b}$, which we express in compact form by
\begin{equation}
\left[\begin{array}{c}
 b_{1}\\b_{2}\\b_{3}\\b_{4}
\end{array}\right]
=\left[\begin{array}{c}
 \Delta b_{1}\\ \Delta b_{2}\\ \Delta b_{3}\\ \Delta b_{4}
\end{array}\right]
+
\frac{1}{M}\left(
\tau e^{i \beta}
\left[\begin{array}{c}
 \sqrt{2 \kappa_{-}}D_{1}\\ \sqrt{2 \kappa_{-}}D_{2}\\ \sqrt{2 \kappa_{+}}D_{3}\\ \sqrt{2 \kappa_{+}}D_{4}
\end{array}\right]\frac{h}{h_{sql}}
\right)
\label{eq:ab}
\end{equation}
%%%%%%%%%%%%Equation Break%%%%%%%%%%%%%%%%%
where
\begin{equation}
\left[\begin{array}{c}
 \Delta b_{1}\\ \Delta b_{2}\\ \Delta b_{3}\\ \Delta b_{4}
\end{array}\right]
=\frac{1}{M}\left(
e^{2 i (\beta+\Phi)}
\left[\begin{array}{cccc}
 C_{11}&C_{12}&C_{13}&C_{14}\\
 C_{21}&C_{22}&C_{23}&C_{24}\\
 C_{31}&C_{32}&C_{33}&C_{34}\\
 C_{41}&C_{42}&C_{43}&C_{44}\\
\end{array}\right]
\left[\begin{array}{c}
 a_{1}\\a_{2}\\a_{3}\\a_{4}
\end{array}\right]\right)
\label{eq:deltab}
\end{equation}
which is analogous to the notation used by Buonanno and Chen \cite{Buonanno2001} with the exception that our definition for the D-coefficients has the term $\sqrt{2 \kappa_{\pm}}$ included in the coefficients rather than factored out which allows us to consider unequal intensities of the two sidebands.  The compact form of our expression uses the definitions
%%%%%%%%%%%%Equation Break%%%%%%%%%%%%%%%%%
\begin{equation}
M\equiv
\left(e^{4 i (\beta +\Phi )} \rho ^2-2 e^{2 i (\beta +\Phi )} \cos (2 \phi
   ) \rho +1\right) \left(e^{4 i (\beta +\Phi )} \rho ^2-e^{2 i (\beta
   +\Phi )} \left(2 \cos (2 \phi )+\sin (2 \phi ) \left(\kappa _-+\kappa
   _+\right)\right) \rho +1\right)
     \end{equation}
%%%%%%%%%%%%Equation Break%%%%%%%%%%%%%%%%%
\begin{multline}
C_{11}=C_{22}\equiv
\frac{1}{2} \Bigg( \left(2
   \rho ^2+\tau ^2\right) \left(e^{4 i (\beta +\Phi )} \rho ^2-2 e^{2 i
   (\beta +\Phi )} \cos (2 \phi ) \rho +1\right) \sin (2 \phi ) \kappa _-
   -2  e^{-2 i (\beta +\Phi )} \times
   \\
   \left(e^{4 i (\beta +\Phi )} \left(\rho ^2+\tau ^2\right) \rho +\rho
   -e^{2 i (\beta +\Phi )} \left(2 \rho ^2+\tau ^2\right) \cos (2 \phi
   )\right) \left(e^{4 i (\beta +\Phi )} \rho ^2-e^{2 i (\beta +\Phi )}
   \left(2 \cos (2 \phi )+\sin (2 \phi ) \kappa _+\right) \rho
   +1\right)\Bigg)
  \end{multline}
  %%%%%%%%%%%%Equation Break%%%%%%%%%%%%%%%%%
\begin{multline}
C_{33}=C_{44}\equiv
-\frac{1}{2} e^{-2 i (\beta +\Phi )} \Bigg((\left(e^{4 i (\beta +\Phi )}
   \rho ^2-2 e^{2 i (\beta +\Phi )} \cos (2 \phi ) \rho +1\right) \times
   \\
    \left(2
   e^{4 i (\beta +\Phi )} \left(\rho ^2+\tau ^2\right) \rho +2 \rho -e^{2 i
   (\beta +\Phi )} \left(2 \rho ^2+\tau ^2\right) \left(2 \cos (2 \phi
   )+\sin (2 \phi ) \kappa _+\right)\right) -
   \\
   2 e^{2 i (\beta +\Phi )} \rho 
   \left(e^{4 i (\beta +\Phi )} \left(\rho ^2+\tau ^2\right) \rho +\rho
   -e^{2 i (\beta +\Phi )} \left(2 \rho ^2+\tau ^2\right) \cos (2 \phi
   )\right) \sin (2 \phi ) \kappa _-\Bigg)  \end{multline}
%%%%%%%%%%%%Equation Break%%%%%%%%%%%%%%%%%
\begin{multline}
C_{12}\equiv-\tau ^2 \sin (\phi ) \Bigg(e^{4 i (\beta +\Phi )} \left(2 \cos (\phi
   )+\sin (\phi ) \kappa _-\right) \rho ^2-2 e^{2 i (\beta +\Phi )} \times
 \\
  \left(\cos (2 \phi ) \left(2 \cos (\phi )+\sin (\phi ) \kappa
   _-\right)+   \cos (\phi ) \sin (2 \phi ) \kappa _+\right) \rho+2 \cos
   (\phi )+\sin (\phi ) \kappa _-\Bigg)
 \end{multline}
%%%%%%%%%%%%Equation Break%%%%%%%%%%%%%%%%%
\begin{multline}
C_{34} \equiv-\tau ^2 \sin (\phi ) \left(\left(e^{4 i (\beta +\Phi )} \rho ^2-2 e^{2 i
   (\beta +\Phi )} \cos (2 \phi ) \rho +1\right) \left(2 \cos (\phi )+\sin
   (\phi ) \kappa _+\right)-4 e^{2 i (\beta +\Phi )} \rho  \cos ^2(\phi )
   \sin (\phi ) \kappa _-\right) \end{multline}
%%%%%%%%%%%%Equation Break%%%%%%%%%%%%%%%%%
\begin{equation}
C_{13}=C_{24}=C_{31}=C_{42} \equiv-\left(e^{4 i (\beta +\Phi )} \rho ^2-1\right) \tau ^2 \cos (\phi ) \sin
   (\phi ) \sqrt{\kappa _- \kappa _+}
   \end{equation}
%%%%%%%%%%%%Equation Break%%%%%%%%%%%%%%%%%
\begin{equation}
C_{14}=C_{32} \equiv-\left(e^{2 i (\beta +\Phi )} \rho  \tau +\tau \right)^2 \sin ^2(\phi )
   \sqrt{\kappa _- \kappa _+}
   \end{equation}
%%%%%%%%%%%%Equation Break%%%%%%%%%%%%%%%%%
 \begin{multline}
C_{21} \equiv-\tau ^2 \cos (\phi ) \Bigg(-\left(e^{4 i (\beta +\Phi )} \rho ^2+1\right)
   \left(2 \sin (\phi )-\cos (\phi ) \kappa _-\right)-2 e^{2 i (\beta +\Phi
   )} \rho  \cos (2 \phi ) \left(\cos (\phi ) \kappa _--2 \sin (\phi
   )\right)+
   \\
   2 e^{2 i (\beta +\Phi )} \rho  \sin (\phi ) \sin (2 \phi )
   \kappa _+\Bigg)
\end{multline}
%%%%%%%%%%%%Equation Break%%%%%%%%%%%%%%%%%
\begin{multline}
C_{43} \equiv-\tau ^2 \cos (\phi ) \left(4 e^{2 i (\beta +\Phi )} \rho  \cos (\phi )
   \kappa _- \sin ^2(\phi )+\left(e^{4 i (\beta +\Phi )} \rho ^2-2 e^{2 i
   (\beta +\Phi )} \cos (2 \phi ) \rho +1\right) \left(\cos (\phi ) \kappa
   _+-2 \sin (\phi )\right)\right)
   \end{multline}
%%%%%%%%%%%%Equation Break%%%%%%%%%%%%%%%%%
\begin{equation}
C_{23}=C_{41} \equiv-4 e^{2 i (\beta +\Phi )} \left(e^{2 i (\beta +\Phi )} \rho -1\right)^2 \tau ^2 \cos ^2(\phi ) \sqrt{\kappa _- \kappa _+}
\end{equation}
%%%%%%%%%%%%Equation Break%%%%%%%%%%%%%%%%%
\begin{equation}
D_{1} \equiv-\left(e^{2 i (\beta +\Phi )} \rho +1\right) \sin (\phi ) \left(e^{2
   i (\beta +\Phi )} \rho   \kappa _+ \sin
   (2 \phi )+\left(e^{4 i (\beta +\Phi )} \rho ^2-e^{2 i
   (\beta +\Phi )} \left(2 \cos (2 \phi )+\sin (2 \phi ) \kappa _+\right)
   \rho +1\right)\right)
   \end{equation}
%%%%%%%%%%%%Equation Break%%%%%%%%%%%%%%%%%
\begin{equation}
D_{2} \equiv- \left(e^{2 i (\beta +\Phi )} \rho -1\right) \cos (\phi ) \left(e^{2 i (\beta +\Phi )} \rho  \kappa _+
   \sin (2 \phi )+\left(e^{4 i (\beta +\Phi )} \rho ^2-e^{2 i (\beta +\Phi )} \left(2 \cos (2 \phi )+\sin (2 \phi ) \kappa
   _+\right) \rho +1\right)\right)
\end{equation}
%%%%%%%%%%%%Equation Break%%%%%%%%%%%%%%%%%
\begin{equation}
D_{3} \equiv- \left(e^{2 i (\beta +\Phi )} \rho +1\right) \sin (\phi ) \left(e^{2 i (\beta +\Phi )} \rho  \kappa _-
   \sin (2 \phi )+\left(e^{4 i (\beta +\Phi )} \rho ^2-e^{2 i (\beta +\Phi )} \left(2 \cos (2 \phi )+\sin (2 \phi ) \kappa _-\right) \rho
   +1\right) \right)
\end{equation}
and
%%%%%%%%%%%%Equation Break%%%%%%%%%%%%%%%%%
\begin{multline}
D_{4} \equiv- \left(e^{2 i (\beta +\Phi )} \rho -1\right) \cos (\phi ) \left(e^{2 i (\beta +\Phi )} \rho  \kappa _- 
   \sin (2 \phi )+\left(e^{4 i (\beta +\Phi )} \rho ^2-e^{2 i (\beta +\Phi )} \left(2 \cos (2 \phi )+\sin (2 \phi ) \kappa _-\right) \rho
   +1\right) \right)
\end{multline}
These input output relations of equation \ref{eq:ab} form the basis for our analysis of the quantum-noise limited sensitivity.  We note that by making the substitutions $\kappa_{-} \to \kappa$ and $\kappa_{+} \to 0$ these expressions reproduce equations 2.20-2.24 in Ref. \cite{Buonanno2001} for a signal-recycled interferometer with a conventional input laser spectrum.  To investigate the quantum-noise limted sensitivity of an interferometer with this readout scheme we must first consider how these output fields are converted to a signal through heterodyne detection and demodulation.

\section{Demodulation}
The output field of the interferometer is composed of two parts, a local oscillator and a signal 
(which includes the quantum noise) and can be written as
\begin{equation}
{E}(t)=L(t)+{S}(t)
\end{equation}
Where
\begin{equation}
L(t) = L e^{i \omega_0 t} + {\rm h.c.}
\end{equation}
where h.c. stands for the hermitian conjugate, and $L$ describes the amplitude and phase of the local oscillator.    We follow the analysis of reference \cite{Buonanno2003} and express the signal as 
\begin{eqnarray}
{S}(t) &\equiv&
%\int_{0}^{+\infty}\frac{d\omega}{2 \pi}
%\left[\,e^{-i \omega t}\,{b}_{\omega}+{\rm h.c.}\right],\nonumber \\
%&=&
\int_{-\Lambda}^{+\Lambda} \frac{d\Omega}{2 \pi}\frac{1}{2\sqrt{2}}
\left[
b_{1}(\Omega) e^{-i (\omega_0-\omega_{m}+\Omega)t}
+i b_{2}(\Omega) e^{-i (\omega_0-\omega_{m}+\Omega)t}
+b_{3}(\Omega) e^{-i (\omega_0+\omega_{m}+\Omega)t}
+i b_{4}(\Omega) e^{-i (\omega_0+\omega_{m}+\Omega)t}
+{\rm h.c.}\right]
\nonumber \\
&+&({\rm contributions\;at\;irrelevant\;frequency\;bands}), 
\end{eqnarray}
where $\Lambda$ is the demodulation bandwidth.  
For simplicity we omit terms that are at frequencies that will not contribute to the final expression.

The detected photocurrent is proportional to the square of the field giving
\begin{equation}
i(t)\propto E_{out}^{2}=L^{2}(t)+2 L(t) S(t) + S^{2}(t)
\end{equation}
of which only the cross term $2 L(t) S(t)$ has significant amplitude at the modulation frequency which is detected by demodulation and low pass filtering.  This modulated component of the photocurrent is
\begin{multline}
i_{mod}(t)\equiv 2 L(t) S(t) = 2  \left(L e^{i \omega_0 t}+ \rm h.c.\right) \times
\\
\int_{-\Lambda}^{+\Lambda} \frac{d\Omega}{2 \pi}\frac{1}{2\sqrt{2}}
\left[
b_{1}(\Omega) e^{-i (\omega_0-\omega_{m}+\Omega)t}
+i b_{2}(\Omega) e^{-i (\omega_0-\omega_{m}+\Omega)t}
+b_{3}(\Omega) e^{-i (\omega_0+\omega_{m}+\Omega)t}
+i b_{4}(\Omega) e^{-i (\omega_0+\omega_{m}+\Omega)t}
+{\rm h.c.}\right]
\end{multline}
The demodulated output signal comes from mixing (i.e. multiplying) the modulated photocurrent with $\cos{(\omega_{m} t +\phi_{d})}$ and low pass filtering with a cut-off frequency of $\Lambda$ giving
\begin{eqnarray}
i_{demod}(t)&=&\frac{1}{T}\int_{T-t}^{T}{i_{mod}(t)\cos{(\omega_{m} t +\phi_{d})} dt}
\nonumber\\ 
&\approx& 
\int_{-\Lambda}^{+\Lambda} \frac{d\Omega}{2 \pi}\frac{L}{2\sqrt{2}}
e^{-i \Omega t}
\left[
b_{1}(\Omega) e^{i \phi_{d}}
+i b_{2}(\Omega) e^{i \phi_{d}}
+b_{3}(\Omega) e^{-i \phi_{d}}
+i b_{4}(\Omega) e^{-i \phi_{d}}
\right]+{\rm h.c.}
\label{eq:doublesided}
\end{eqnarray}
where $T=1/\Lambda$ is the time constant associated with the low pass filter and we approximate the low pass filter as  being a perfect step-function filter passing only frequencies within its bandwidth.
Using the relations
\begin{eqnarray}
b_{1}(-\Omega)&=&b_{1}(\Omega)\nonumber
\\
b_{2}(-\Omega)&=&b^{\dagger}_{2}(\Omega)
\\
b_{3}(-\Omega)&=&b_{3}(\Omega)\nonumber
\\
b_{4}(-\Omega)&=&b^{\dagger}_{4}(\Omega)
\end{eqnarray}
we can rewrite the double-sided frequency integral in equation \ref{eq:doublesided} as the single-sided integral
\begin{eqnarray}
i_{demod}(t)
&\approx& 
\int_{0}^{+\Lambda} \frac{d\Omega}{2 \pi}\frac{L}{\sqrt{2}}
e^{-i \Omega t}
\left[
b_{1}(\Omega) e^{i \phi_{d}}
+i b_{2}(\Omega) e^{i \phi_{d}}
+ b_{3}(\Omega) e^{-i \phi_{d}}
+i b_{4}(\Omega) e^{-i \phi_{d}}
\right]+{\rm h.c.}
\end{eqnarray}
which in the frequency domain is
\begin{eqnarray}
i_{demod}(\Omega)
&\approx& 
\frac{L}{\sqrt{2}}
\left[
b_{1}(\Omega) e^{i \phi_{d}}
+i b_{2}(\Omega) e^{i \phi_{d}}
+ b_{3}(\Omega) e^{-i \phi_{d}}
+i b_{4}(\Omega) e^{-i \phi_{d}}
\right]
\end{eqnarray}
Setting $\zeta_{\pm}=\zeta_{0}\mp \phi_{d}-\pi/2$
where $\zeta_{0}\equiv\arg{(L)}$
we can express this as
\begin{eqnarray}
i_{demod}(\Omega)
&\approx& 
\frac{|L|}{\sqrt{2}}
\left[
i b_{1}(\Omega) e^{i \zeta_{-}}
- b_{2}(\Omega) e^{i \zeta_{-}}
+i  b_{3}(\Omega) e^{-i \zeta_{+}}
- b_{4}(\Omega) e^{-i \zeta_{+}}
\right]
\end{eqnarray}
and with the definitions
\begin{eqnarray}
b_{\zeta+}&=&b_{3}\sin{\zeta_{+}}+b_{4}\cos{\zeta_{+}}
\\
b_{\zeta-}&=&b_{1}\sin{\zeta_{-}}+b_{2}\cos{\zeta_{-}}\nonumber
\label{eq:bzeta}
\end{eqnarray}
which represent the signal due to the output field in a band centered at  $\omega_{0}\pm\omega_{m}$ in the $\zeta_{\pm}$ quadratures respectively, the demodulated signal is
\begin{equation}
i_{demod}(\Omega)
\approx
-\frac{|L|}{\sqrt{2}}
\left(
b_{\zeta-}+b_{\zeta+}
\right).
\end{equation}
The values of $b_{i}$ in expression \ref{eq:bzeta} are given by the input-output relations of equations \ref{eq:ab} and \ref{eq:deltab}.  Since the output fields $b_{\zeta\pm}$ each can contain vacuum fluctuations and signals from gravitational waves, the quantum-noise limited output is  unlike that of  equation 13 of reference [\cite{Buonanno2003}] for heterodyne readout of an interferometer with a conventional input spectrum for which the frequency regions around $\omega_{0}\pm2 \omega_{m}$ contribute noise (called non-stationary shot-noise) but not signal.  Thus we have shown that {\em heterodyne detection with a suppressed carrier input spectrum is not subject to non-stationary shot-noise}.
 
\section{Noise spectral density and sensitivity}
To compute the noise spectral density we express the magnitude of the noise by the equivalent signal level $h_{n}$.  From equation \ref{eq:ab} we have
\begin{equation}
h_{n}=h_{sql}\left|
\frac{
\Delta b_{1} \sin{\zeta_{-}}+\Delta b_{2} \cos{\zeta_{-}}+\Delta b_{3} \sin{\zeta_{+}}+\Delta b_{4} \cos{\zeta_{+}}
}
{
\tau \left(\sqrt{2 \kappa_{-}} D_{1} \sin{\zeta_{-}} +\sqrt{2 \kappa_{-}} D_{2} \cos{\zeta_{-}}+\sqrt{2 \kappa_{+}} D_{3} \sin{\zeta_{+}}+\sqrt{2 \kappa_{+}} D_{4} \cos{\zeta_{+}}\right)
}\right|.
\end{equation}
We find the noise spectral density by
 Equation.~(22) 
of Ref. \cite{KLMTV}
\begin{equation}
\frac{1}{2}\,2\pi\,\delta(\Omega - \Omega^\prime)\,S^\zeta_h(f) 
=  \frac{1}{2}
\langle {\rm in} | h_n(\Omega)\,h_n^\dagger(\Omega^\prime) + 
h_n^\dagger(\Omega^\prime)\,h_n(\Omega)|{\rm in} \rangle.
\end{equation}
and assuming the input field a is in the vacuum state such that
\begin{equation}
\langle 0_a| a_i\,a^\dagger_{j^\prime} |0_a \rangle +\langle 0_a| a^\dagger_i\,a_{j^\prime} |0_a \rangle = 
\frac{1}{2}\,2\pi\,\delta(\Omega- \Omega^\prime)\,
\delta_{i j}\,,
\end{equation} 
the noise spectral density is 
\begin{equation}
S=\frac{|A_{1}|^{2}+|A_{2}|^{2}+|A_{3}|^{2}+|A_{4}|^{2}}{|
\tau \sqrt{2 \kappa_{-} (D_{1} \sin{\zeta_{-}} +D_{2} \cos{\zeta_{-}} )+2 \kappa_{+} (D_{3} \sin{\zeta_{+}} +D_{2} \cos{\zeta_{+}} )}
|^{2}} h_{sql}^{2}.
\end{equation}
The $A_{i}$ terms in the numerator represent the noise contributions from each of the 4 input `a' fields and are defined as
\begin{eqnarray}
A_{i}&\equiv&C_{1i}\sin{\zeta_{-}}+C_{2i}\sin{\zeta_{-}}+C_{3i}\sin{\zeta_{+}}+C_{4i}\sin{\zeta_{+}}
\end{eqnarray}
We then define the sensitivity as the square root of the noise spectral density.  For reference we reference this to the sensitivity associated with the standard quantum limit, $h_{sql}$

\subsection{Balanced laser sidebands}
\begin{figure}[!h]
\includegraphics[width=3 in]{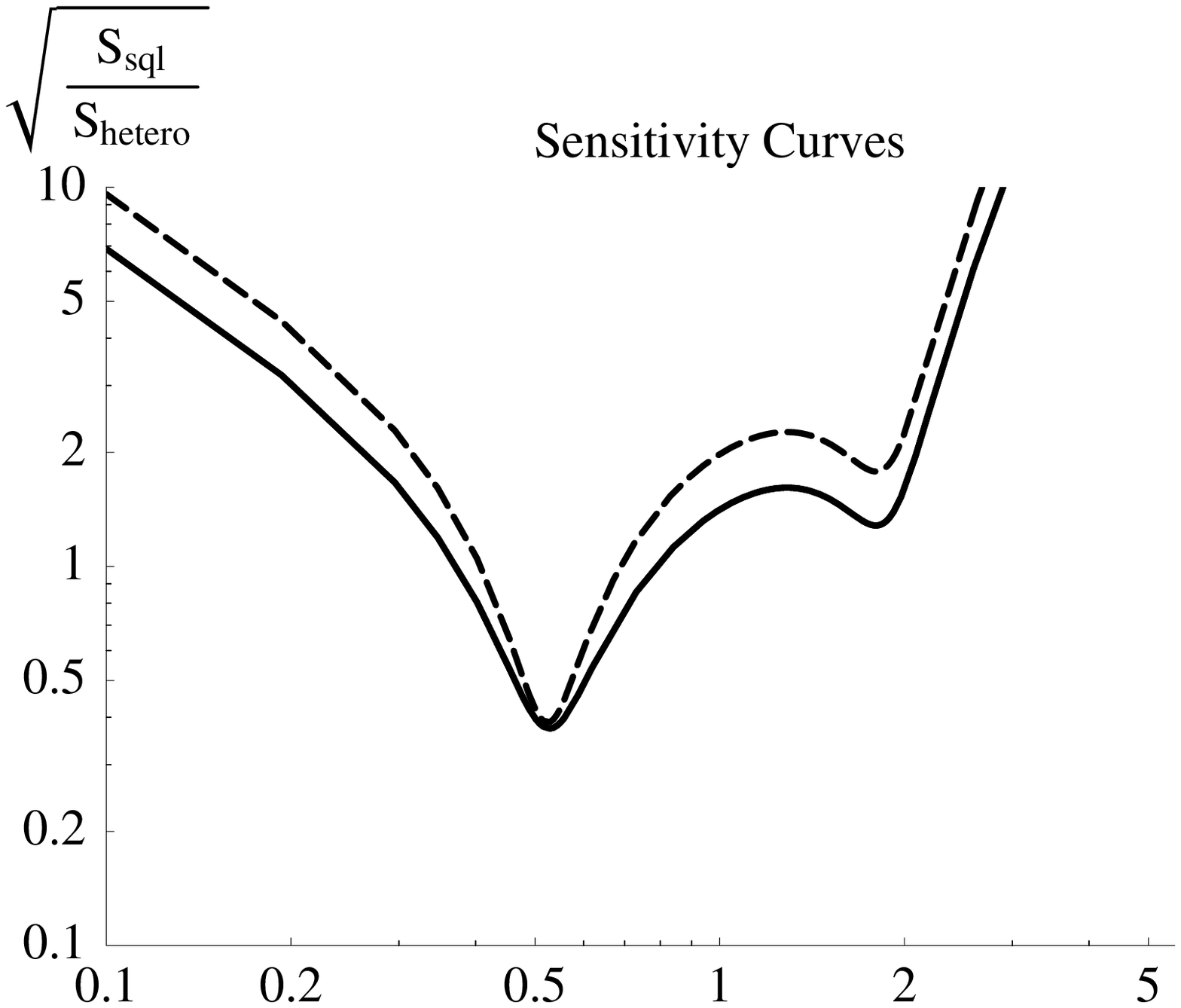}
\includegraphics[width=3 in]{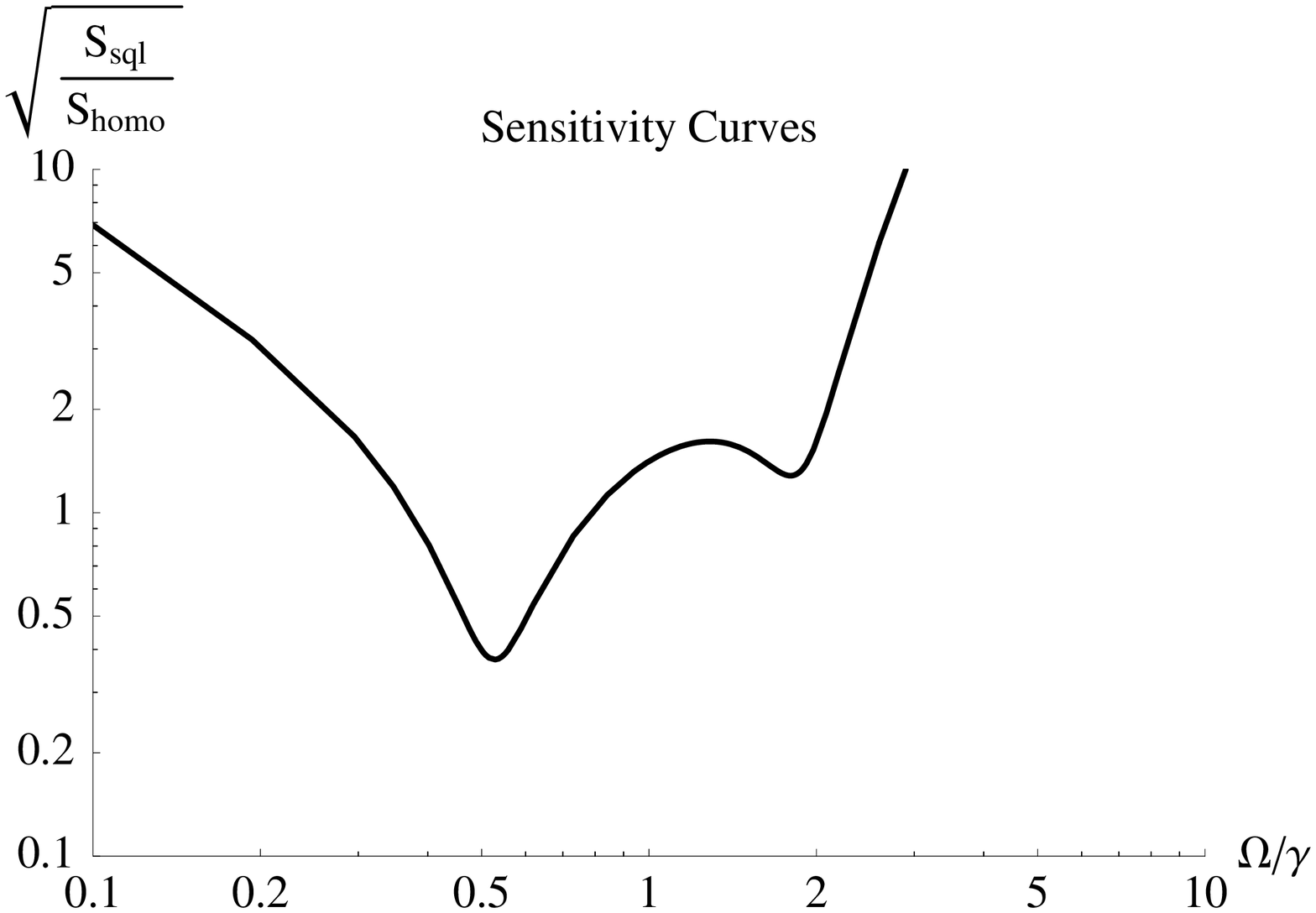}
\caption{\label{fig:S} The left hand figure shows the sensitivity for heterodyne readout with a conventional input spectrum (dashed line) and with the carrier-suppressed input spectrum (solid line).  The sensitivity with the carrier-suppressed input spectrum exactly matches that achievable with homodyne detection  (shown in the right hand figure).   For the carrier-suppressed input spectrum  $I_{-}=I_{+}=I_{sql}/2$, while the conventional spectrum used for heterodyne and homodyne readout has  $I_{0}=I_{sql}$.  For all spectra shown $\rho=0.9$, $\tau=\sqrt{1-\rho^{2}}$, $\Omega=\gamma/2$, $\phi=\pi/2-0.47$, $\zeta_{0}=0$ and $\phi_{d}=0$.}
\end{figure}

\begin{figure}[!h]
\includegraphics[width=3 in]{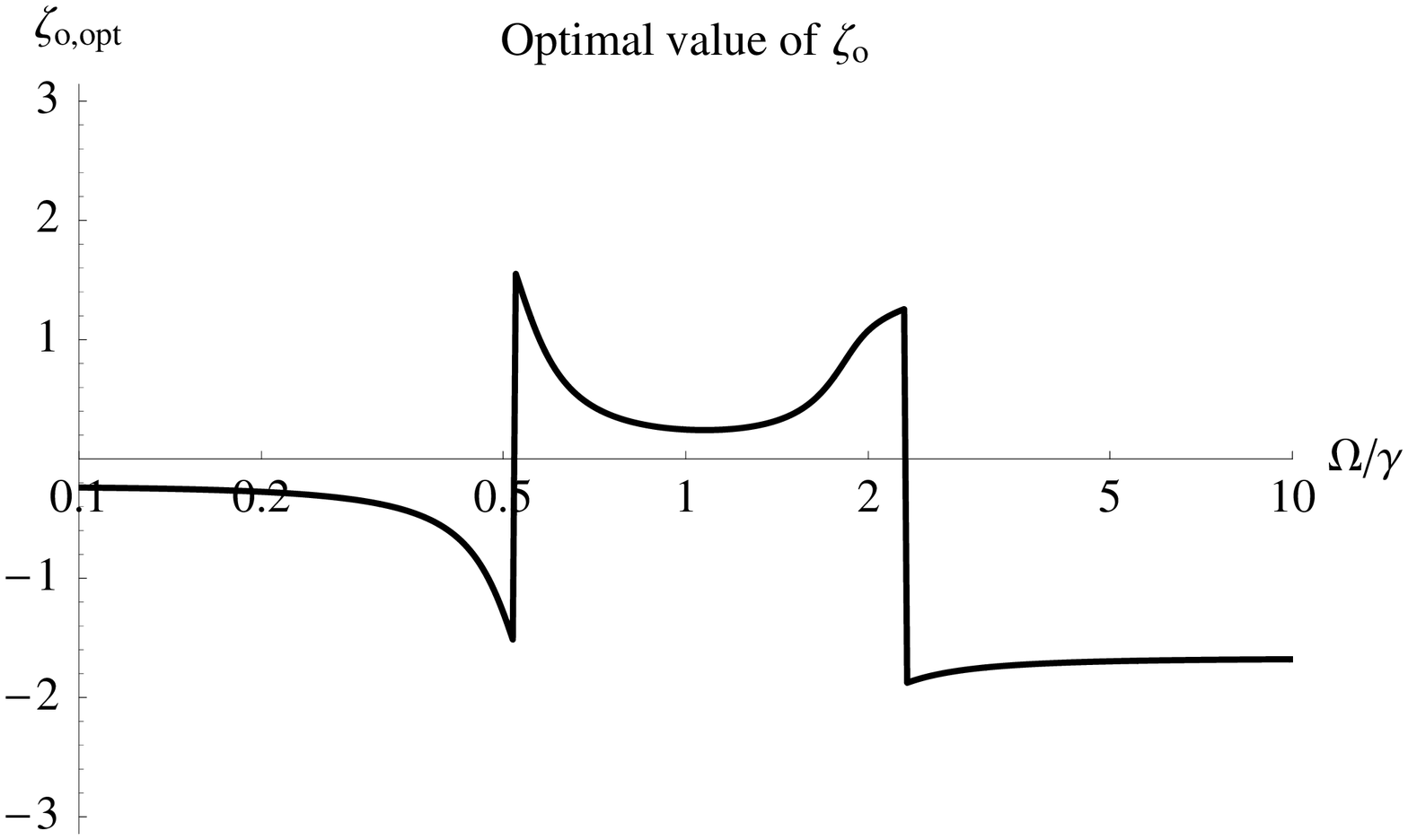}
\includegraphics[width=3 in]{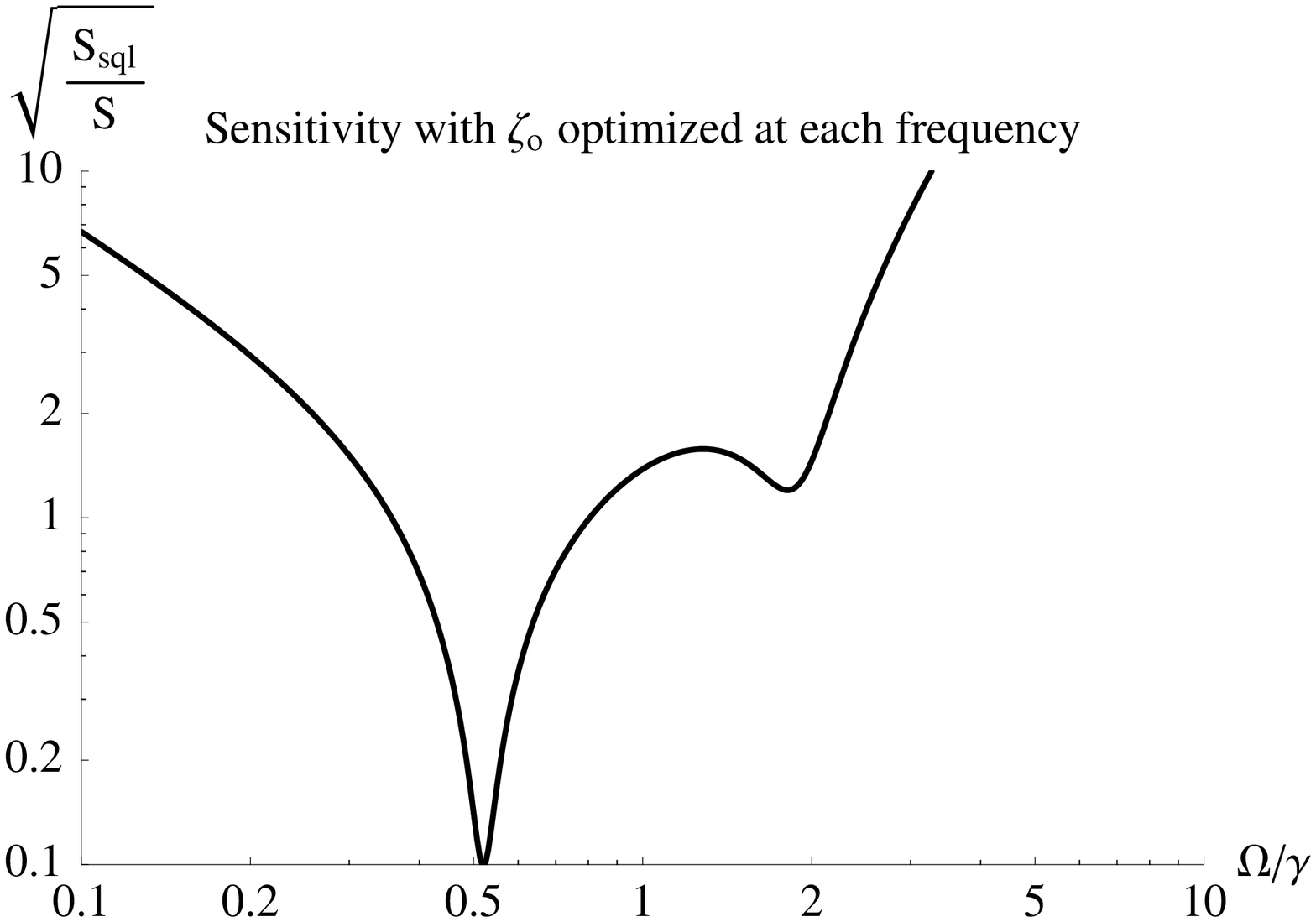}
\caption{\label{fig:zetavsfreq} Left: the optimal value of the local oscillator phase, $\zeta_{0}$ as a function of frequency for an interfereomter with   $\rho=0.9$, $\tau=\sqrt{1-\rho^{2}}$, $I_{-}=I_{+}=I_{sql}/2$, $\Omega=\gamma/2$, $\phi=\pi/2-0.47$, and $\phi_{d}=0$. Right: the sensitivity at each frequency with the optimal local oscillation phase.}
\end{figure}

We now consider the special case of balanced laser sidebands such that $ \kappa_{-}=\kappa_{+}$.  With this assumption $C_{11}=C_{22}=C_{33}=C_{44}$, $C_{12}=C_{34}$, $C_{13}=C_{24}=C_{31}=C_{42}$, $C_{14}=C_{32}$, $C_{21}=C_{43}$, $C_{23}=C_{41}$
$D_{1}=D_{3}$ and $D_{2}=D_{4}$.  With these simplifications, the noise spectral density is an even function of the demodulation phase $\phi_{d}$ and is minimized for $\phi_{d} \to 0$.  The noise spectral density is plotted in figure \ref{fig:S} and compared to the noise spectral density with homodyne readout and with heterodyne readout using a conventional input spectrum.  The noise spectral density for this heterodyne readout with a  carrier-suppressed input spectrum is identical to that with a homodyne readout scheme and is quieter than that for heterodyne readout with a conventional input spectrum.
The optimal value for $\zeta_{0}$ the phase of the local oscillator is frequency dependent.  The optimal sensitivity as a function of frequency and the local oscillator phase necessary for that sensitivity are shown in figure \ref{fig:zetavsfreq}

\subsection{Unbalanced laser sidebands}
\begin{figure}[!hp]
\includegraphics[width=3 in]{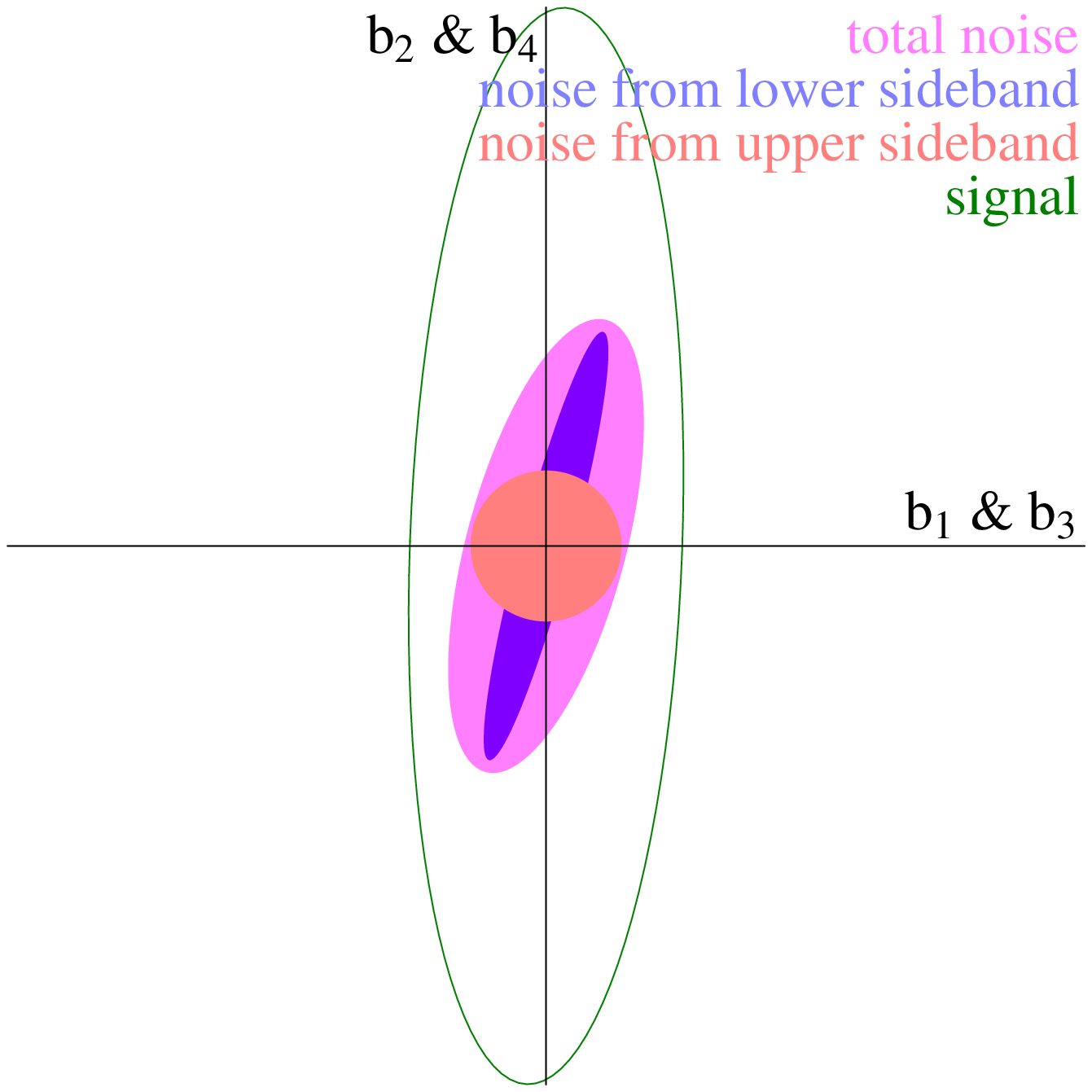}
\includegraphics[width=3 in]{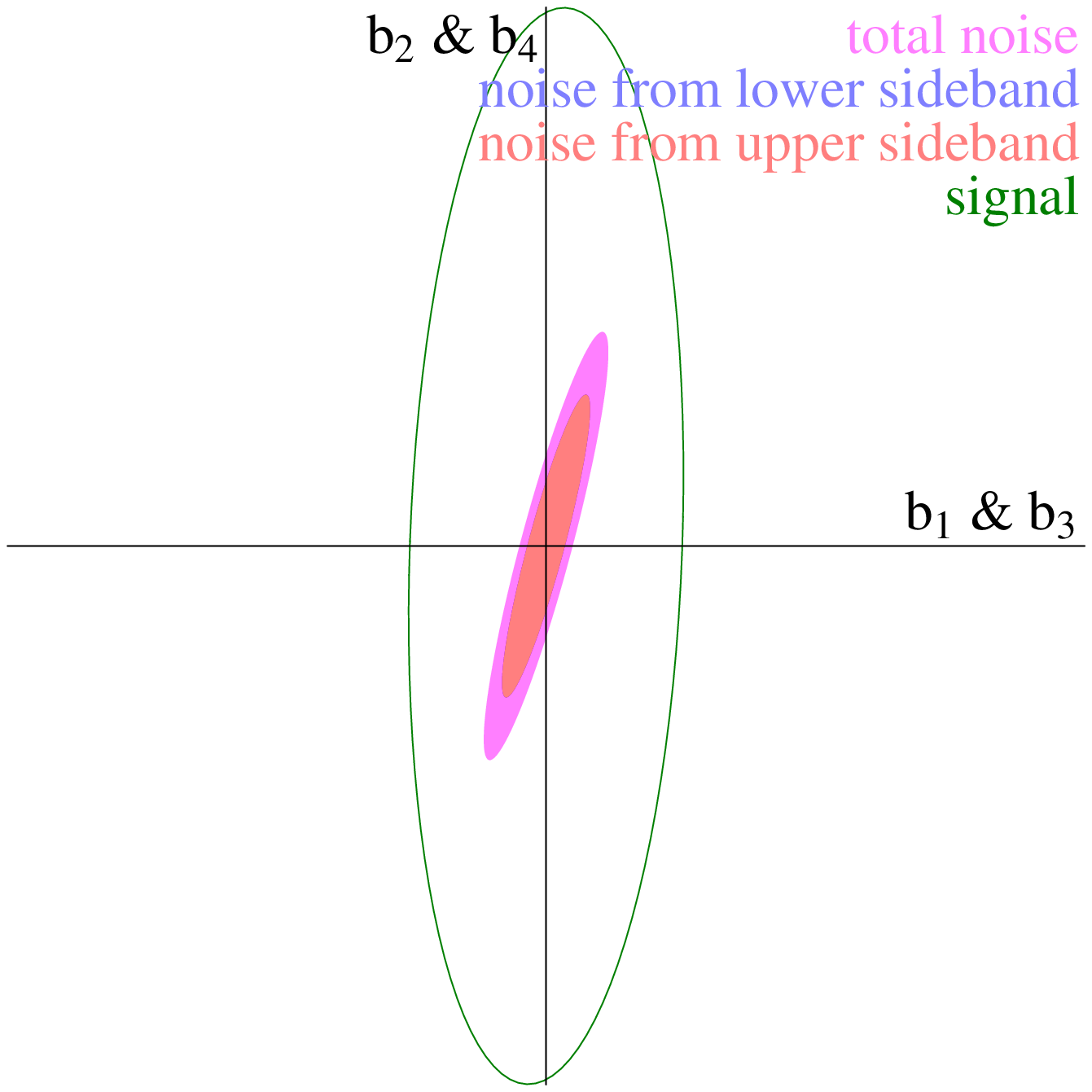}

\caption{\label{fig:noiseellipseunbalanced} The total quantum noise ellipse (outer shaded ellipse) and its components from frequencies near the lower sideband (dark shaded inner ellipse) and the upper sideband (light shaded inner ellipse). The ellipse representing a signal of magnitude $h_{sql}$ is shown for comparison.  On the left is the noise ellipse for unbalanced ($I_{-}=I_{sql}$, $I_{+}=0$) sidebands on the right  is the noise ellipse for balanced sidebands ($I_{-}=I_{+}=I_{sql}/2$).  For both cases  $\rho=0.9$, $\tau=\sqrt{1-\rho^{2}}$, $\Omega=\gamma/2$, $\phi=\pi/2-0.47$, and $\phi_{d}=0$.   If homodyne detection were used the noise ellipse would be equivalent to the contribution from the lower frequency sideband in the unbalanced case (inner dark ellipse on the left) and the ellipse would be identical the total noise ellipse for balanced detection (outer ellipse on the right).}
\end{figure}

In the general case of unbalanced laser sidebands, the lack of symmetry causes the optimal demodulation phase to differ from zero in a frequency dependent manner,  however, unlike the phase of the local oscillator which is fixed, several different demodulation phases can be used to readout the signal in parallel channels, allowing for a noise spectral density that approximates that which would be achieved by having the optimal demodulation phase at each frequency.  The noise spectral density of the readout with unbalanced sidebands and frequency optimized demodulation angle is always between the best case of balanced sidebands and the worst case of totally unbalanced sidebands.  The quantum noise ellipse \cite{Caves1985} is shown in figure \ref{fig:noiseellipseunbalanced} for the totally unbalanced case and the balanced case.  

\begin{figure}[!h]
\includegraphics[width=3 in]{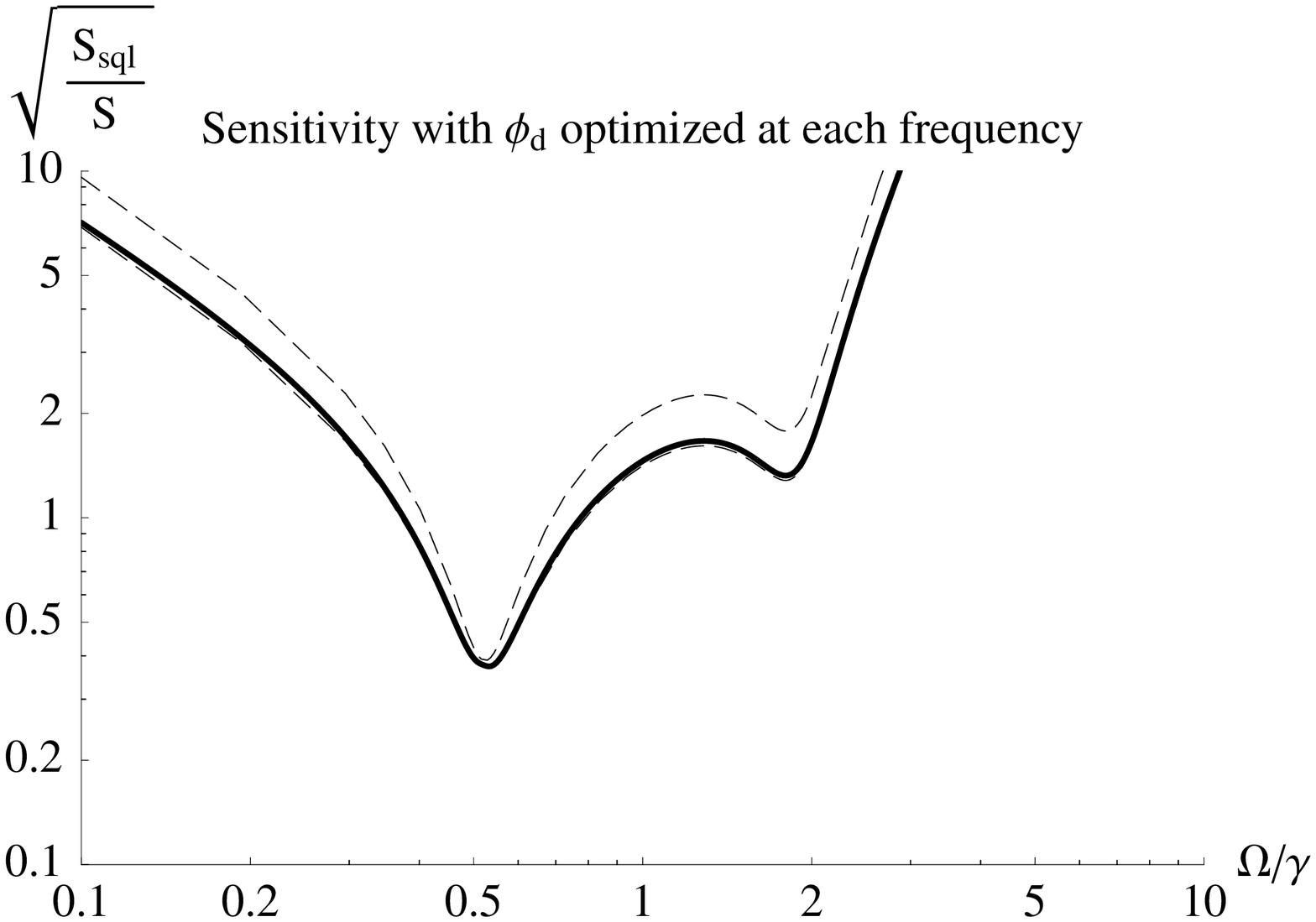}
\includegraphics[width=3 in]{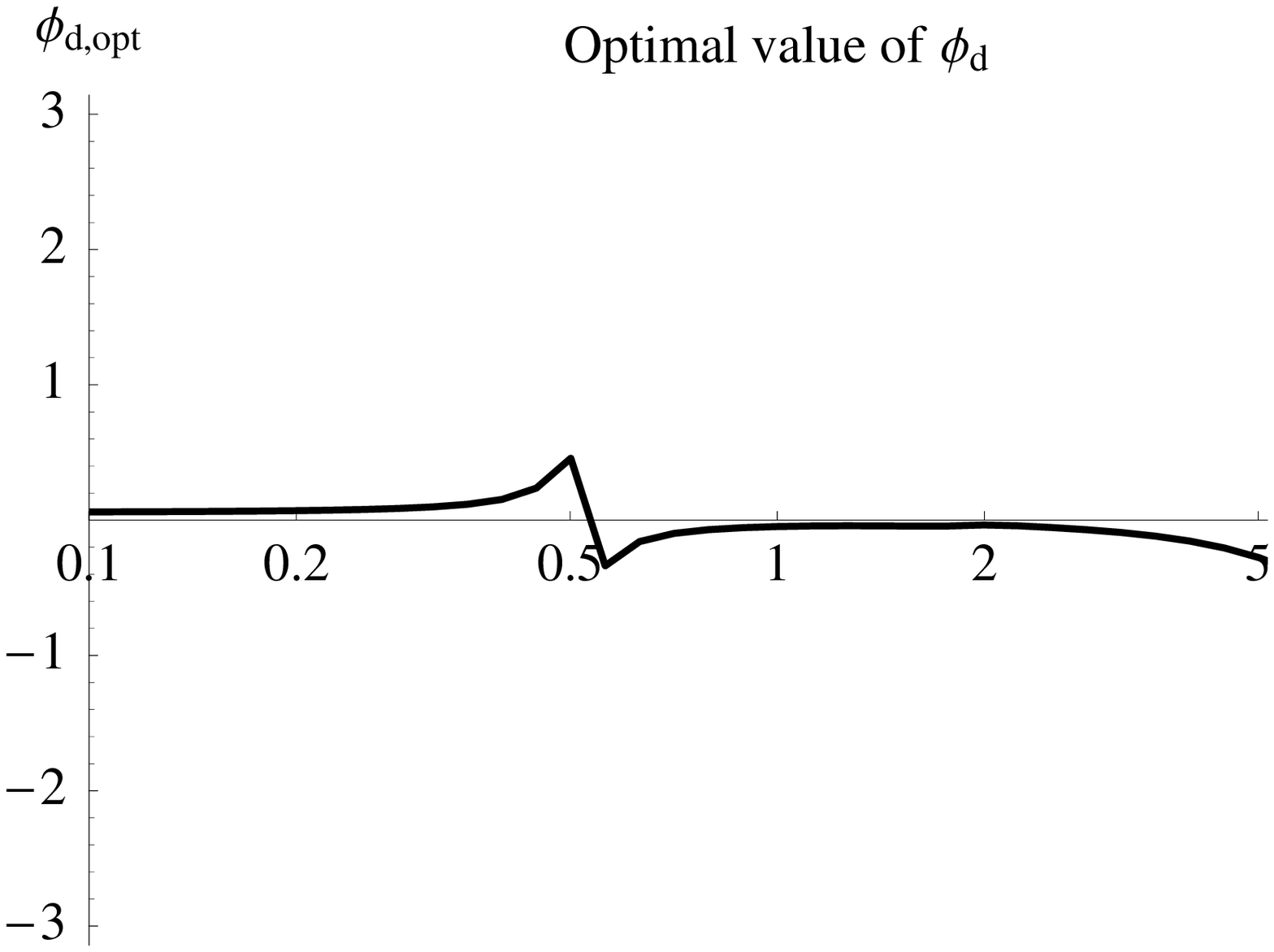}
\caption{\label{fig:unbalancedsens} Left: The sensitivity with unbalanced sidebands ($I_{-}=0.25 I_{sql}$, $I_{+}=0.75 I_{sql}$) is plotted as the solid line.  It is bounded by the sensitivity with balanced sidebands ($I_{-}=0.5 I_{sql}$, $I_{+}=0.5 I_{sql}$) on the bottom (lower dashed line) and the sensitivity with completely imbalanced sidebands ($I_{-}=0 I_{sql}$, $I_{+}=I_{sql}$) on the top (upper dashed line).  For all curves  $\rho=0.9$, $\tau=\sqrt{1-\rho^{2}}$, $\phi=\pi/2-0.47$, and $\zeta_{0}=0$.  For the completely balanced and completely imbalanced sideband cases $\phi_{d}=0$, while for the unbalanced case the optimal demodulation phase, shown in the right panel is assumed at each frequency. }
\end{figure}

It may seem that the ability to generate an output spectrum with a frequency optimized demodulation angle is a benefit of having unbalanced sidebands, however 
the optimal demodulation angle with balanced sidebands is frequency independent, so there is no disadvantage to balanced sidebands

\section{Conclusion}
Future gravitational wave interferometric detectors will be limited by quantum-noise over much of the detection band.  The details of how the readout scheme affects the quantum-noise limited sensitivity are an important consideration for the detector design .  With the conventional input spectrum the quantum-noise with heterodyne readout is worse than with homodyne readout due to the presence of non-stationary shot noise -- vacuum fluctuations at twice the modulation frequency which are down converted into the output signal by mixing with the modulation sidebands.   With the suppressed-carrier input spectrum presented here the quantum-noise limited sensitivity with heterodyne readout is equivalent to that with homodyne readout because the non-stationary shot noise is not present.  Although both heterodyne readout schemes and homodyne readout schemes have other strengths and weaknesses, we show that the use of heterodyne detection does not necessarily lead to an increase in the quantum-limited noise floor.

\begin{acknowledgments}
This work was supported by the National Science Foundation grant PHY-0140297,``The Stanford Advanced Gravitational Wave Detector Research Program''
\end{acknowledgments}

\end{document}